\documentclass[twocolumn]{aastex631}
\usepackage{amsmath}
\usepackage{xspace}

\usepackage{multirow,url}

\newcommand{\cxo}{{\it Chandra}}
\newcommand{\xmm}{{\it XMM-Newton}\xspace}

\newcommand{\revI}[1]{{#1}}
\newcommand{\revII}[1]{{#1}}

\begin{document}

\title{Hyperluminous Supersoft X-Ray Sources in the {\em Chandra} Catalog}

\correspondingauthor{Andrea Sacchi}\email{andrea.sacchi@cfa.harvard.edu}
\author[0000-0002-7295-5661]{Andrea Sacchi}
\author{Kevin Paggeot}
\affiliation{Center for Astrophysics $\vert$ Harvard \& Smithsonian, 60 Garden Street, Cambridge, MA 20138, USA}
\author[0000-0002-4773-1463]{Steven Dillmann}
\affiliation{Institute for Computational and Mathematical Engineering, Stanford University, Stanford, CA 94305, USA}
\author[0000-0002-5069-0324]{Juan Rafael Mart\'inez-Galarza}
\affiliation{Center for Astrophysics $\vert$ Harvard \& Smithsonian, 60 Garden Street, Cambridge, MA 20138, USA}
\affiliation{AstroAI, Center for Astrophysics $\vert$ Harvard \& Smithsonian, 60 Garden Street, Cambridge, MA 20138, USA}
\author[0000-0003-4511-8427]{Peter Kosec}
\affiliation{Center for Astrophysics $\vert$ Harvard \& Smithsonian, 60 Garden Street, Cambridge, MA 20138, USA}







\begin{abstract}
Hyperluminous supersoft X-ray sources, such as bright extragalactic sources characterized by particularly soft X-ray spectra, offer a unique opportunity to study accretion onto supermassive black holes in extreme conditions. Examples of hyperluminous supersoft sources are tidal disruption events, systems exhibiting quasi-periodic eruptions, changing-look AGN, and anomalous nuclear transients.
Although these objects are rare phenomena amongst the population of X-ray sources, we developed an efficient algorithm to identify promising candidates exploiting archival observations. 
In this work, we present the results of a search for hyperluminous supersoft X-ray sources in the recently released {\em Chandra} catalog of serendipitous X-ray sources. This archival search has been performed via both a manual implementation of the algorithm we developed and a novel machine-learning-based approach. 
This search identified a new tidal disruption event, which might have occurred in an intermediate-mass black hole. This event occurred between 2001 and 2002, making it one of the first tidal disruption events ever observed by {\em Chandra}.

\end{abstract}



\section{Introduction} \label{sec:intro}

We define hyperluminous supersoft X-ray sources (HSSs) as bright extragalactic sources of X-rays characterized by a supersoft spectrum. Their X-ray luminosity exceeds $10^{41}$~erg/s, and their spectra are extremely steep. When modeled with a power law profile the typical photon indices are $\Gamma\gtrsim3$, if a black body model is adopted, typical temperatures are $kT\lesssim100$~eV, and with very few exceptions, all of their emission is below 2~keV. HSSs are usually spatially coincident with the nuclear region of their host galaxies. This, coupled with their high luminosities and steep spectra, indicates that the origin of their emission is accretion onto supermassive black holes (SMBHs). 

The best-studied and best-represented subclass of HSSs is that of tidal disruption events (TDEs). TDEs occur when a star orbit happens to plunge deep enough into the potential well of a SMBH to be disrupted by tidal forces. In the standard scenario, roughly half of the resultant stellar debris remains on bound orbits and eventually falls back on the SMBHs and later is accreted after forming a bright accretion disc. The electromagnetic emission generated by this newly formed accretion disc is predicted to peak in the ultraviolet (UV) band, while the steep Wien tail of the spectrum would fall in the X-ray band. This emission is expected to rapidly rise and then slowly decay over timescales of days to months. These phenomena were first theorized \citep{hills75,rees88,phinney89}, and first observed by ROSAT \citep{bade96}. The properties of the first TDEs observed were exactly the ones predicted: bright and soft X-ray transients from otherwise quiescent galactic nuclei. The importance of these phenomena was immediately clear. They represent unique opportunities to study newly-formed accretion flows on otherwise quiescent and relatively ``light'' SMBHs ($\sim10^6$~M$_\odot$). This is because the tidal radius for a Sun-like star is smaller than the Schwarzschild radius for a SMBH with a mass $\gtrsim10^8$~M$_\odot$, hence a star would not get disrupted, but rather directly swallowed by any more massive SMBH: one such encounter would not generate a TDE. 

After the first TDEs were discovered in the X-ray band, these phenomena were found to be bright across the full electromagnetic spectrum, and today we know of about a hundred of TDEs observed in every wavelength, from radio \citep{cendes24}, to infrared \citep{masterson24}, optical/UV \citep{gezari12,vanVelzen12,hammerstein23}, X-ray \citep{saxton20,sazonov21}, and Gamma-ray \citep{burrows11}. The simple picture described above struggles to explain all of the data accumulated on these phenomena in the last three decades. Ongoing debates revolve around where, when, and how the multi-wavelength properties of TDEs originate and are correlated \citep[][e.g.]{dai18,bonnerot21}. To further complicate this scenario, new classes of bright and supersoft transients from SMBHs have emerged in recent years. The discovery of quasi-periodic eruptions (QPEs), spectacular X-ray flashes from galactic nuclei recurring over hourly time scales\citep{miniutti19,giustini20,arcodia21,quentin23,nicholl24}, repeated/partial TDEs \citep{payne21,wevers23,malyali23,lin24,guolo24}, long-lived TDEs \citep{lin17a,he21,goodwin22} and active galactic nuclei (AGN) in a purely thermal state \citep{terashima12,lin17b}, calls for further theoretical and observational efforts in order to explain the nature and characteristic of these phenomena. The properties of this plethora of sources are similar enough to those of ``classical'' TDEs that a comprehensive model of the latter should also be able to somewhat account for the former. Although all of these classes of sources exhibit different temporal evolutions, their X-ray properties, in terms of spectrum and overall luminosity, are strikingly similar; hence we collect them all and define the HSS category.

The path to advance our understanding of these phenomena, along with further modeling and theoretical investigations, is to identify new HSSs. Today's standard way of detecting these sources is to exploit their extreme variability, and repeated scans of the sky, performed at different wavelengths, have proven to be highly effective in this task. However, no such scan is currently being performed in the X-ray band, and overlooked HSSs may still be awaiting identification in archival datasets, particularly in available catalogs. In a recent publication \citep{sacchi23} (hereafter Sa23), we scanned the \xmm\ source catalog \citep{webb20} for HSSs. Rather than relying on the variability of the sources, we exploited the unique combination of high luminosity and supersoft spectral properties of HSSs to identify them in the catalog, which allows us to include in our search sources for which only single-epoch observations are available. The results described in Sa23 are promising: we successfully identified nine previously unknown HSSs, amongst which four are candidate TDEs. A follow-up observation performed with \xmm\ confirmed the nature of one of these candidates, making it the first ever X-ray TDE identified solely through X-ray spectroscopy.

In this paper, we repeat this process on the recently-released updated version of the \cxo\ source catalog. This time, alongside the algorithm employed to search the \xmm\ source catalog, we also adopt an unsupervised machine-learning (ML) method by \cite{dillmann24} that relies on learning a physically meaningful representation of the individual X-ray photon events for each catalog source. This accomplishes multiple goals at once. We successfully identify a new TDE, we validate the ML algorithm and explore its potentiality for future applicability to other catalogs as well as other types of sources.

This paper is organized as follows: in Section \ref{sec:classic} we describe how we identified HSSs in the \cxo\ source catalog via our ``classic'' algorithm; in Section \ref{sec:source} we discuss the properties of the new TDE we identified; in Section \ref{sec:ml} we present the comparison of our algorithm with the ML one; and in Section \ref{sec:conc} we draw our conclusions.

\section{Classical approach}\label{sec:classic}

\subsection{Sample selection}

\begin{figure*}[t]
\centering
\includegraphics[width = 0.7\textwidth]{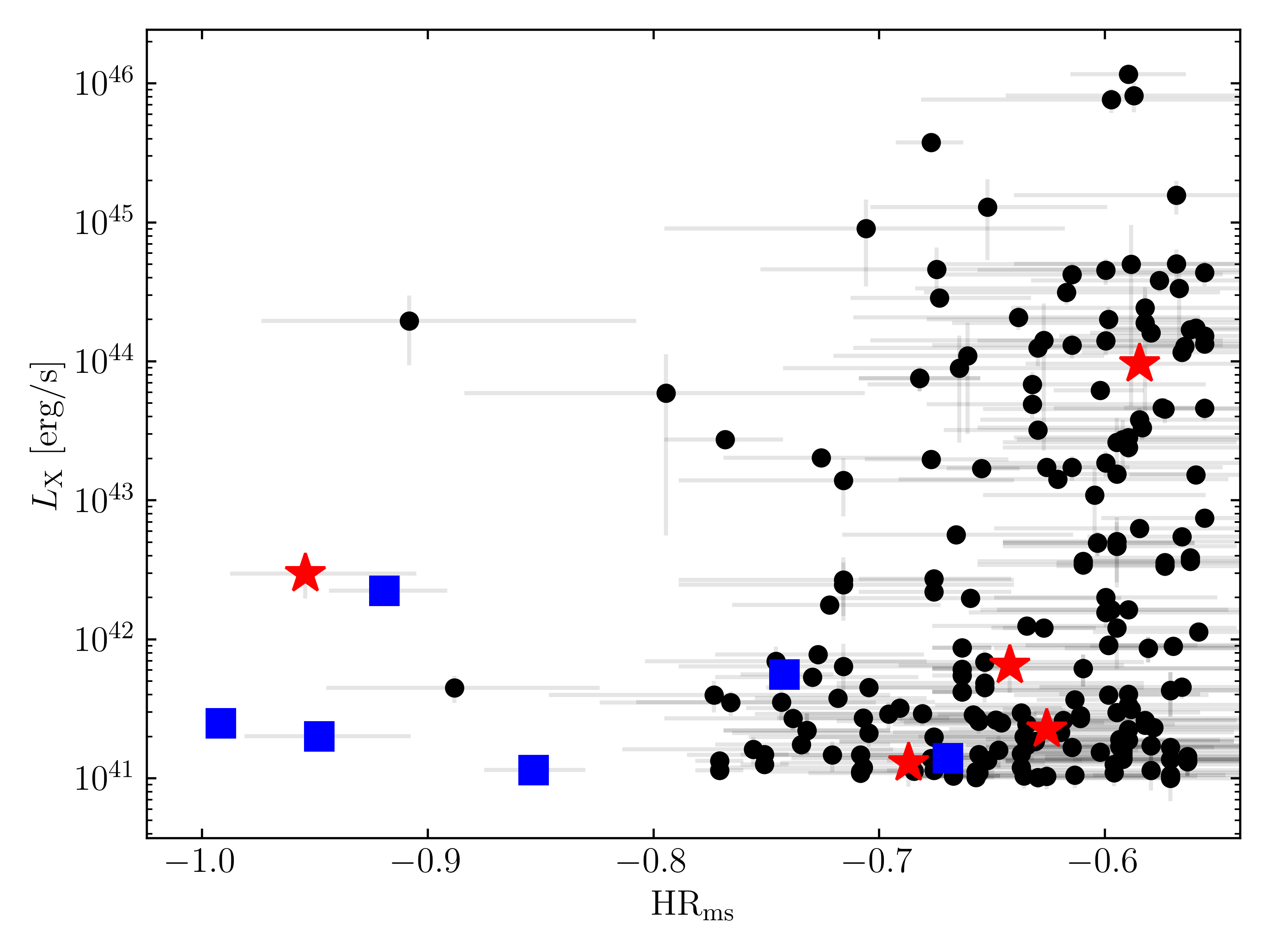}
\caption{X-ray luminosity as a function of the HR in the soft-to-medium band for the source selected via their X-ray properties. In black are the sources we then excluded based on their optical classification, in blue are the HSSs known in the literature, and in red are the sources classified as non-active galaxies whose X-ray spectrum we finally analyzed. \revI{The error bars correspond to $1\sigma$ uncertainties.} \label{fig:gxlx}}
\end{figure*}

\begin{table*}
\centering
\caption{Properties of the five sources we selected as candidate HSSs (indicated by red stars in Fig. \ref{fig:gxlx}). Identifier and redshift pertain to the optical counterpart from SIMBAD. Coordinates, HR, and luminosity (computed from the broad 0.5--7~keV band flux) refer to the X-ray source. The reported classification is the one decided upon the analysis of the X-ray spectrum of each source.}
\label{tab:src}
        \begin{tabular}{l|ccccc|c}
        ID & R.A. & DEC. & $z$ & HR$_\textup{ms}$ & $L_\textup{X}$~[erg/s] & classification\\
        \hline
        2XMMi J215039.5--055335 & 327.6646 & --5.8934 & 0.392 & --0.58 & $9.59\times10^{44}$ & AGN\\
        IC 1792 & 34.7544 & 34.4623 & 0.035 & --0.63 & $2.26\times10^{41}$ & collisionally-ionized gas\\
        2MASX J23272553+2635075 & 351.8564 & 26.5854 & 0.058 & --0.69 & $1.29\times10^{41}$ & collisionally-ionized gas\\
        ESO 152--37 & 29.1705 & --55.9271 & 0.091 & --0.64 & $6.51\times10^{41}$ & collisionally-ionized gas \\
        SDSS J103557.83+572500.2 & 158.9909 & 57.4167 & 0.102 & --0.95 & $2.96\times10^{42}$ & TDE\\
        \hline
    \end{tabular}
\end{table*}

Our process for identifying new HSSs in the \cxo\ source catalog (CSC 2.1, \citealt{evans24}), was similar to the process described in Sa23 for the \xmm\ source catalog. We proceeded to download the full catalog, which amounts to more than 1.3 million individual detections of about 407,000 unique sources observed before the end of 2021. We accessed the catalog through the application which represents its main interface\footnote{The CSC 2.1 application is available at \url{http://cda.cfa.harvard.edu/cscview/}}. We used individual detections rather than stacked datasets of single sources. This is because HSSs might exhibit extreme variability, and different observations of the same source can have significantly different properties.

For all detections, we retrieved their fluxes, hardness ratios (HRs), significances \revI{(Ss)} in different energy bands, and errors in each of these quantities. \revI{We adopted aperture flux to be as independent as possible from particular choices of spectral models, and flux significances, which is a simple estimate of the ratio of the flux measurement to its average error.} We selected our sources based on their S in the soft (0.5--1.2~keV) band, HR in the soft to medium (1.2--2~keV) band, and luminosity in the broad (0.5--7~keV) band. We compute 
the \revI{absorbed} luminosities from the X-ray fluxes using redshift measurements. We crossmatched the \cxo\ catalog with all the redshift catalogs available in SIMBAD \citep{simbad00}. We performed the crossmatch using the tool TOPCAT \citep{topcat05}, with a matching radius corresponding to the error on the source position \revI{(provide by the \cxo\ catalog at 90\% confidence level)}, \revI{and considering the best optical match for each X-ray source, as multiple X-ray sources might be hosted in the same galaxy.} Knowing the redshift, the luminosity values are computed by assuming a standard flat $\Lambda$CDM cosmology with $\Omega_\textup{M}=0.3$ and $H_0=70$ km s$^{-1}$ Mpc$^{-1}$. \revI{We note that, given that the emission from HSSs is extremely soft, the natural choice would be to employ also the ultra-soft (0.2--0.5~keV) band. However, the progressive and dramatic decrease in effective area of \cxo's detectors, especially at ultra-soft energies \citep{plucinsky18,plucinsky22},  implies that the ultra-soft band is not reliable for source selection.}

To build our sample we discarded all sources with a broad-band luminosity $L_\textup{X}<10^{41}$~erg/s, HR~$>-0.55$ in the soft-to-medium band and S~$>7$ in the soft band. The choice of HR corresponds to a photon index $\Gamma\gtrsim3$ if the X-ray spectra of the sources were modeled with a power-law profile. These three criteria ensure that we are excluding sources powered by stellar processes (which rarely exceed the chosen luminosity threshold), the bulk of the AGN population (which usually exhibits much flatter spectra), and \revI{that the source statistics is high enough} to perform a reliable spectral and timing analysis. This resulted in 222 detections from 125 individual sources. Figure \ref{fig:gxlx} shows the X-ray luminosity as a function of the HR for all selected detections.

These 222 detections represent about four times as many HSS candidates as were present in the XMM catalog from Sa23, even though the initial catalogs are similarly-sized. This is because we are not imposing any conditions on the ultra-soft (0.2--0.5~keV) band, and hence we expect more spurious sources to be present in this sample. To further clean our selection we thus exploit the optical classification of the sources, which we retrieved along with the redshift information from SIMBAD. Straightaway we identified and excluded 24 spurious sources \revI{(86 detections)}: ten sources associated with stars, one located behind a supernova remnant, four associated with a foreground galaxy, seven whose soft X-ray photons are associated with the emission from galaxy clusters, and two whose emission is extended and attributable to a pair of interacting galaxies. Similarly to what we found in Sa23, at this stage, the bulk of our sample is composed of sources optically classified as AGN, 75 \revI{(96 detections)} as broad-line AGN, and 15 \revI{(24 detections)} Seyfert 2 galaxies (mostly classified by \citealt{veron-cetty10} and \citealt{paris14}). The X-ray emission of these sources passes our supersoft criterion as it is characterized by a strong soft excess dominating the energy range below 2~keV. However, they also exhibit hard X-ray emission, which can be well-reproduced by a power-law model (seen through different ranges of absorption material). Overall, the X-ray emission of these sources is compatible with their optical classification and we will not address them further. 

This cleaning process leaves us with a sample of 11 candidate sources. Six of these are well-known HSSs. RX~J1301.9+2747 and GSN~069 are peculiar AGN showing QPEs \citep{miniutti19,giustini20}. WINGS~J134849.88+263557.5 and ASASSN-14li are two famous TDEs, the former discovered in the galaxy cluster Abell~1795 \citep{maksym13}, and the latter often referred to as a ``textbook'' TDE given that its multi-wavelength emission follows precisely all the predictions for these phenomena \citep{maksym14}. SDSS~J150052.07+015453.8 is a long-lived TDE discovered by \citet{lin17a} and [FWB2009]~HLX-1 is a hyperluminous X-ray source (HLX) hosted in ESO 243-49 which is thought to be powered by accretion over a bonafide intermediate-mass black hole (IMBH) \citep{farrell09}. These sources are exactly the target of our search and testify to the reliability of our algorithm and cleaning procedure.

This leaves five sources, which are classified as non-active galaxies based on their optical emission. The properties of these five sources are listed in Tab. \ref{tab:src}. To investigate their nature further, we performed a full X-ray spectral analysis of all the publicly available datasets for each source.

\subsection{X-ray spectral analysis}

All \cxo\ datasets were reprocessed and reduced with the Chandra Interactive Analysis of Observations software package (\texttt{CIAO}, v.4.12; \citealt{fruscione06}) and the \texttt{CALDB} 4.9.0 release of the calibration files. \revI{For each source, spectra,} spectral redistribution matrices, and ancillary response files were generated using the \texttt{CIAO} script \texttt{specextract}. Spectra were extracted from regions encompassing a point-spread function (PSF) fraction of about 95\%. Local backgrounds were estimated from source-free annular regions, centered on the source position. \revI{These annular regions have inner radii of $\approx10$~arcsec and outer radii of $\approx30$~arcsec. If these regions are too close to the detector borders, a circular region of $\approx20$~arcsec radius was adopted instead.}


One source, 2XMMi~J215039.5$-$055335 has been observed by \cxo\ twice. In both observations, the X-ray spectrum of the source is well modeled by a power-law with a photon index $\Gamma\approx2$ when the full 0.5--7~keV energy band is considered. This suggests that the X-ray emission from this source is attributable to a typical X-ray corona and, coupled with its high X-ray luminosity ($\approx10^{44}$~erg/s) implies that it is a misclassified AGN. Three sources, IC~1792, 2MASX~J23272553+2635075, and ESO~152$-$37, exhibit X-ray spectra that can be well-reproduced by an \texttt{apec} model, suggesting that the origin of their emission is powered by collisionally-ionized gas, heated up when falling in the gravitational potential well of the galaxy itself. This is also compatible with their luminosity which barely exceeds the $10^{41}$~erg/s threshold. The X-ray spectra and spectral parameters of these sources are reported in Appendix \ref{app:xray_spec}. However, although passing our cuts, these sources are not HSSs: they are either not powered by accretion of gas onto a massive BH, or their X-ray spectrum is not supersoft when the full X-ray energy band is inspected.  

Finally, one source, SDSS~J103557.83+572500.2, exhibits a supersoft thermal X-ray spectrum coupled with a luminosity exceeding $10^{42}$~erg/s. This source is a bonafide HSS and represents a strong candidate TDE. Its multi-wavelength emission is described at length in the next Section.

\section{A new ``old'' TDE}\label{sec:source}

SDSS~J103557.83+572500.2 (J1035+57 hereafter) is a galaxy at spectroscopic redshift $z=0.10203$ \citep{ahumada20}, with a stellar mass of $6.45\times10^{10}$~M$_\odot$ and a star formation rate (SFR) of 4.24 M$_\odot$/y \citep{chang15}, and is cataloged as a starforming galaxy by \citet{toba14}. The infrared colors ($W1-W2=0.258$, from AllWISE, reported by \citealt{kim23}), as well as the optical spectroscopy from the Sloan Digital Sky Survey (SDSS, \citealt{ahumada20}), suggest no ongoing nuclear activity. The X-ray luminosity from unresolved X-ray binaries, based on the reported values of stellar mass and SFR, amounts to $L_{\rm X,XRB}\approx1.3\times10^{40}$~erg/s, estimated employing \citet{lehmer10} relation. From the stellar mass of J1035+57, exploiting the $M_\textup{BH}-M_\star$ relation for non-active galaxies \citep{reines15}, we infer a SMBH mass of $4.8_{-3.2}^{+9.4}\times10^{8}$~M$_\odot$.

\begin{figure}[t]
\centering
\includegraphics[width=0.5\textwidth]{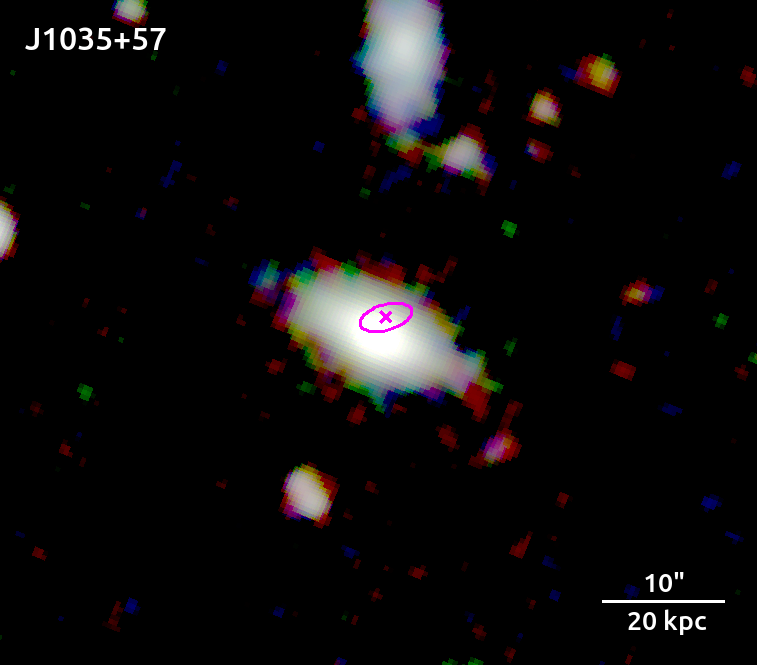}
\caption{Three-color composite image if J1035+57 obtained with \textit{SDSS} \textit{gri}-filters. The cross and the magenta ellipse indicate the centroid and positional uncertainty of the X-ray source detected by \cxo. \label{fig:opt_img}}
\end{figure}

J1035+57, due to its proximity to Lockman Hole, is detected by \cxo\ during a serendipitous visit on 30 April 2002 (ObsID 3346, P.I. Barger). The effective exposure time of this visit is 38.2~ks and the source position lies on the edge of the detector, at about 16' of off-axis angle. Considering this, the X-ray source position is uncertain at about 2", and it is marginally compatible with its host galaxy center. Figure \ref{fig:opt_img} shows an optical image of J1035+57 with the superposed location and uncertainty of the X-ray source.

\begin{figure}[t]
\centering
\includegraphics[width=0.5\textwidth]{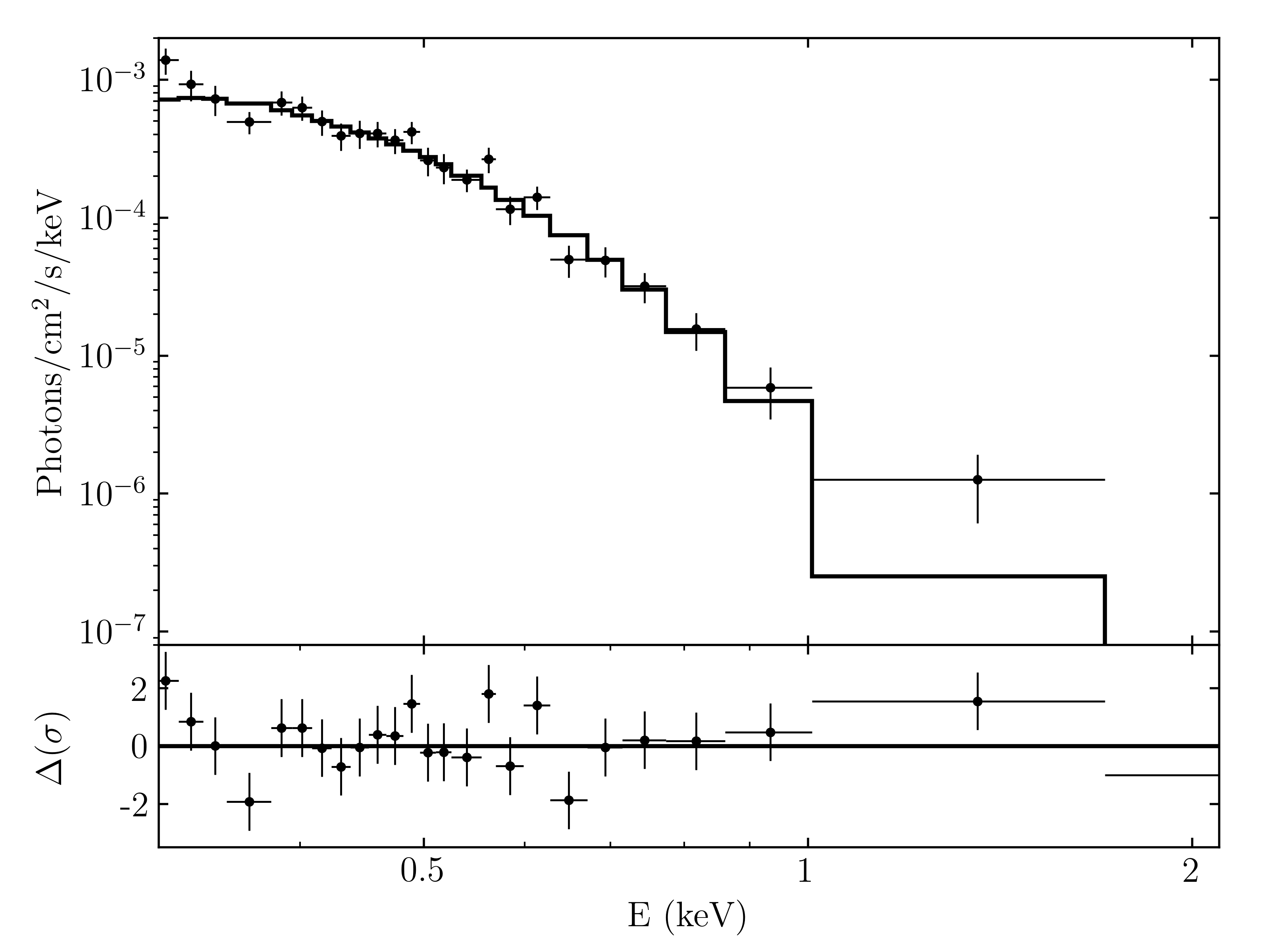}
\caption{X-ray spectrum (upper panels) and residuals (lower panels) of J1035+57. The solid lines show the best-fitting models described in the text. \revII{Only the 0.3--2~keV energy range is shown as above 2~keV the spectrum is background dominated.} \label{fig:xspec}}
\end{figure}

The X-ray spectrum of J1035+57, reported in Fig. \ref{fig:xspec}, can be well reproduced by a simple black-body model (redshifted at the source distance, \texttt{zbbody} in \texttt{XSPEC}). Given the high signal-to-noise ratio of the detection, we rebinned the spectrum to have at least 20 counts in each bin and adopted the standard $\chi^2$ statistic. Considering the supersoft nature of the source's spectrum and the fact that the dataset was acquired in 2002 when \cxo\ effective area in the soft X-ray band was not degraded by molecular contamination, we also include the counts collected in the 0.3--0.5~keV band in our analysis. To the black-body model, we added a layer of neutral absorber (\texttt{TBabs}) with $N_\textup{H}$ fixed to the Galactic value $5.46\times10^{19}$ cm$^{-2}$ \citep{hi4pi16}, using abundances from \citet{wilms00}, with the photoelectric absorption cross-sections from \citet{verner96}. We obtained a satisfactory fit ($\chi^2/\nu=36.3/29\approx1.24$, where $\chi^2$ is the value of the statistic and $\nu$ the degrees of freedom) for a temperature of $85\pm3$~eV and an unabsorbed luminosity of $1.28\pm0.07\times10^{42}$~erg/s in the 0.5--2~keV band. The spectrum residuals offer hints of excess counts in the 1--2~keV band, but adding a power law to reproduce this excess does not significantly improve the fit's quality. Likewise, the quality of the fit is not improved by adding a free-to-vary layer of neutral absorption. The 3-$\sigma$ upper limit on the intrinsic absorption at the source location (or within the host galaxy) is $N_\textup{H,intrinsic}<3\times10^{20}$~cm$^{-2}$. We also attempted fitting the spectrum with \revI{alternative models, such as} a power-law and an \texttt{apec} model, and in both cases, we obtained unacceptable fits ($\chi^2/\nu=57.5/29\approx1.98$ and $\chi^2/\nu=127.9/29\approx4.41$ respectively).

\begin{figure*}[t]
\centering
\includegraphics[width = 0.75\textwidth]{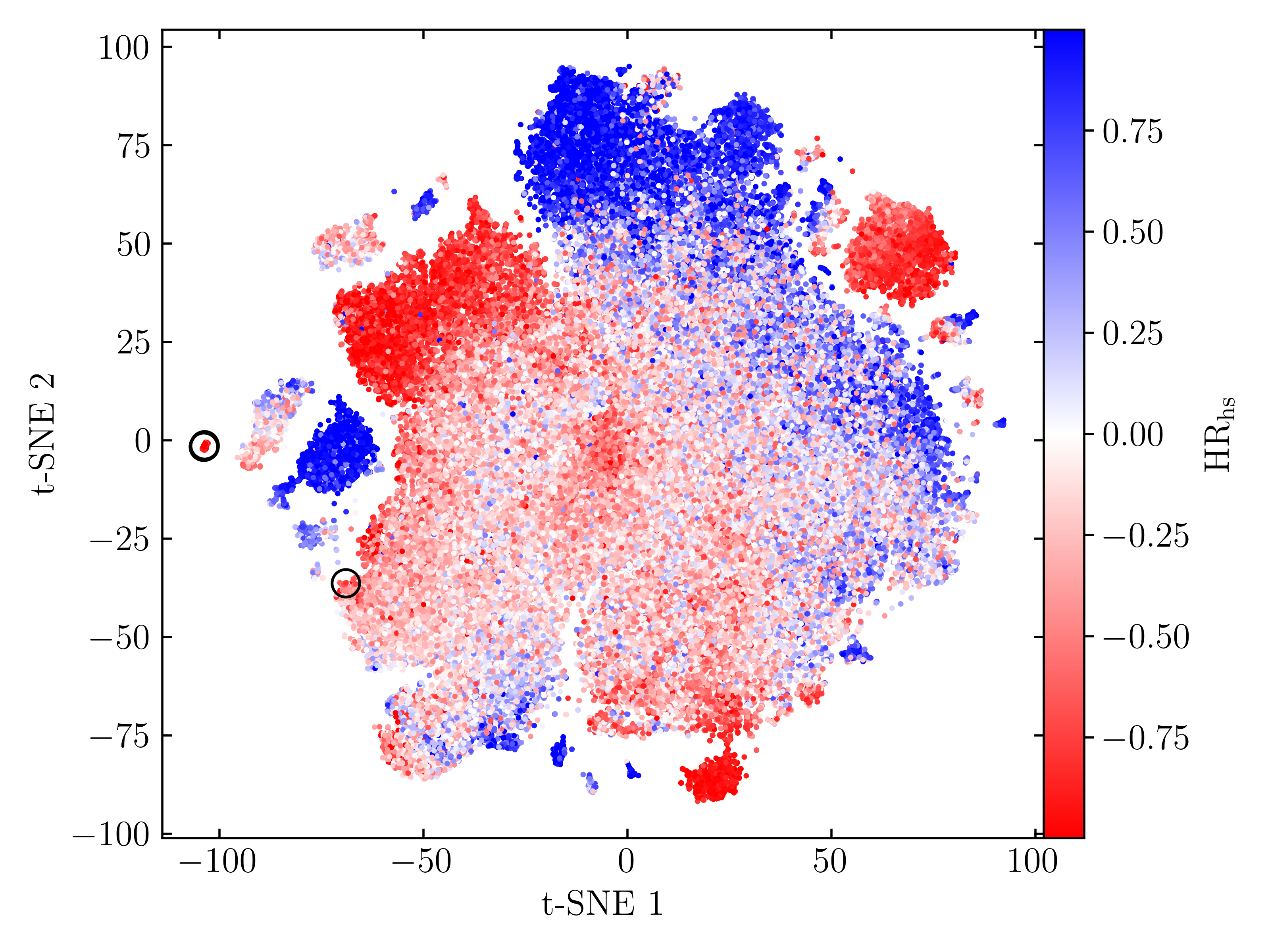}
\caption{t-SNE visualization of the autoencoder embedding representations for the sources considered in \citet{dillmann24}, color-coded by HR$_\textup{hs}$. The black empty circles indicate the location of known HSSs. \label{fig:tsne}}
\end{figure*}

We inspected the X-ray light curve of J1035+57 in different energy bands and detected no significant short-term variability.

The location of J1035+57 fell in the field of view of \cxo\ one year before the reported detection, on May 16th, 2001 (ObsID 1697, P.I. Mushotzky), for a total exposure time of 43.7~ks, and in this case too its position is about 16' off-axis. In this visit the source was not detected and the 3-$\sigma$ upper-limit we computed with the \texttt{CIAO} script \texttt{srcflux} amounts to $1.46\times10^{41}$~erg/s, a value a factor of about nine lower than the detection one year later. No other X-ray telescope visited the source location prior to or since and the Rosat All Sky Survey has not detected J1035+57, although the inferred flux upper limit is too high to be meaningful. 

At the time of the 2002 X-ray detection, the source location was not covered in any other band (optical or infrared) and has not shown significant variability in any observation since. The source is also not detected in any radio band. 

The host galaxy properties, X-ray source location, long-term variability, spectral shape, and luminosity, all suggest that J1035+57 emission is due to a TDE. The non-detection dating one year before the detection places the date of the TDE in the second half of 2001 or the first half of 2002.

We note that the SMBH mass we inferred from the galaxy stellar mass is unconventionally heavy for a TDE, in fact, the Schwarzschild radius of a $4.8\times10^{8}$~M$_\odot$ SMBH is larger than its tidal radius for a Sun-like star, and the Wien tail of multi-color disc spectrum would not fall in the soft X-ray band, but rather on the far ultraviolet band. This opens two possible scenarios: either the inferred SMBH mass is heavily overestimated, making J1035+57 an outlier of the $M_\textup{BH}-M_\star$ relation, or this TDE occurred on an IMBH, hosted in the same galaxy.

\section{Machine-learning approach}\label{sec:ml}


In the previous sections, we manually applied the algorithm developed in Sa23 and successfully identified a previously-overlooked TDE in the \cxo\ catalog. In this section, we explore the applicability and efficiency of a ML technique developed in \cite{dillmann24} for this task. 

In ML, representation learning \citep{bengio13}, refers to the process of training a deep neural network to learn a low-dimensional informative representation of the input data that can be used for various downstream tasks, such as classification or regression on relevant parameters. It can also be used \revI{for similarity searches} and anomaly detection in large datasets characterized by high dimensionality and represents a more flexible and scalable approach with respect to classical methods. 


In the X-ray band, the application of classification algorithms based on ML techniques has only been recently explored \citep{yang22ml,perez24}. Most efforts have focused on the identification of variability patterns, and in particular of fast X-ray transients \citep{dillmann24, dillmann_2025_14589318}. Representation learning algorithms build compressed, low-dimensional representations of high-dimensionality objects. In these learned embedding spaces, anomalies are typically isolated from the bulk of ``normal'' objects, and HSSs, the targets of our search, owing to their spectral properties, are expected to be anomalies in this representation space.

We adopt the framework developed by \citet{dillmann24}, which learns a representation of X-ray sources by training a sparse autoencoder to reconstruct the 2D distribution of arrival times and energies of the photons associated with those sources. After training, the learned representations encode meaningful physical information about the sources, such as variability and spectral properties, that are directly learned from the data rather than computed using relatively complicated pipelines.

This approach, which led to the discovery of the new fast extragalactic transient XRT\,200515 \citep{dillmann24}, was trained using 95,473 \cxo\ filtered event files corresponding to 58,932 sources. This is a smaller dataset than the one we analyzed in the previous section, but notably, it included three HSSs: SDSS~J150052.07+015453.8, WINGS~J134849.88+263557.5, and two observations of ASASSN-14li. \revI{We investigate if the learned representations of these known TDEs are isolated in the representation space with respect to the rest of the X-ray sources. If so, this would indicate the algorithm's ability to identify new HSS candidates.}

\begin{figure}[t]
\centering
\includegraphics[width = 0.5\textwidth]{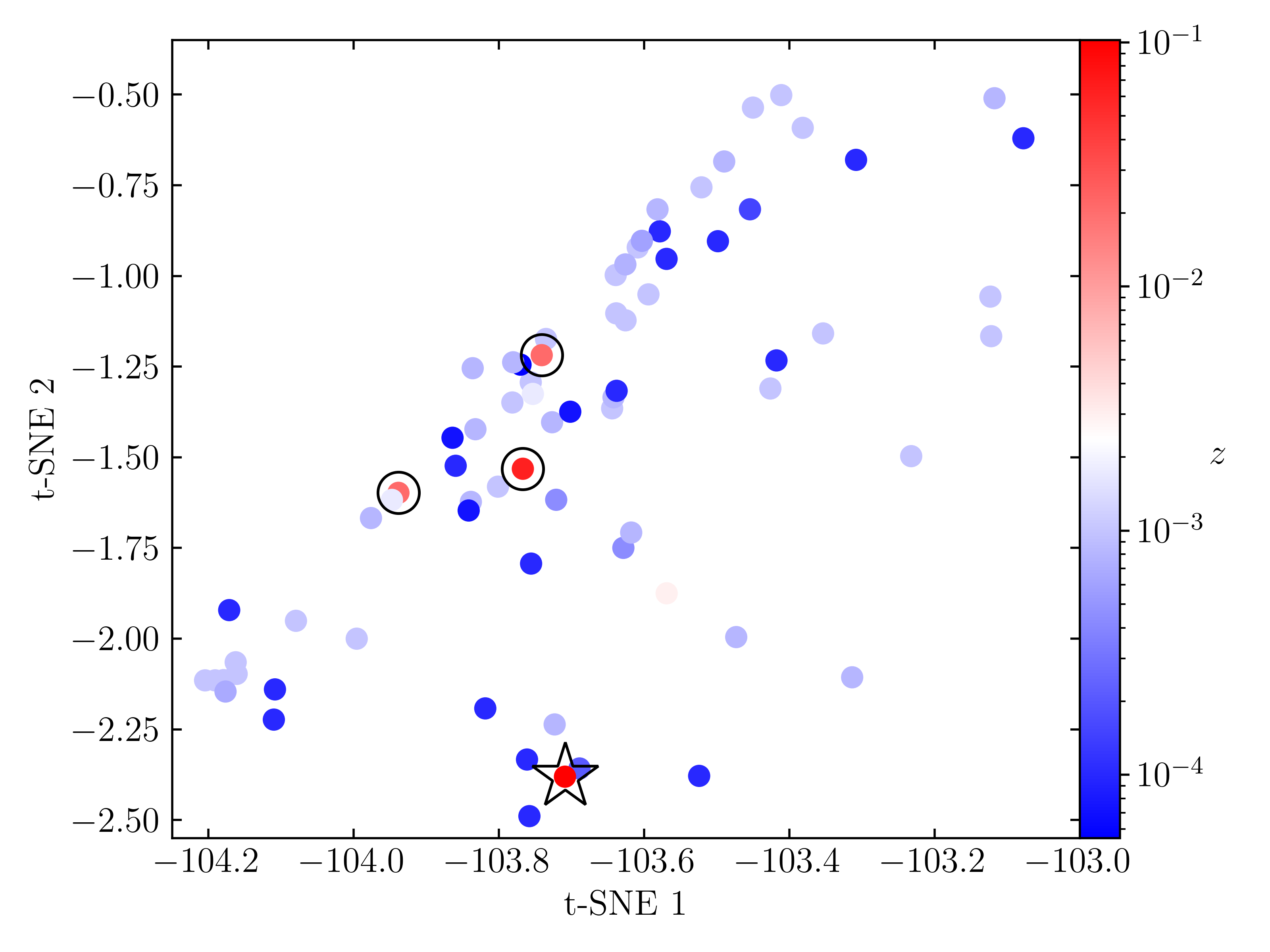}
\caption{Zoom-in of the previous embedding representations, this time color-coded by redshift for. The black empty circles indicate the location of known HSSs. The black empty star indicates the location of the newly identified TDE. \label{fig:tsne_zoom}}
\end{figure}

Figure \ref{fig:tsne} shows a visualization of the autoencoder latent space after training. In addition to the autoencoder, the high-dimensional input data (a vector representing the two-dimensional distribution of photon arrival times and energies), is further reduced to the shown 2D representation using a t-distributed stochastic neighbor embedding (t-SNE, \citealt{maaten08}). The embeddings are color-coded by spectral hardness, and we have highlighted the embedding locations of the known HSSs with the black circles. Three out of the four HSSs' observations, \revI{the two of ASASSN-14li and the single of WINGS~J134849.88+263557.5,} cluster in the same ``island'' of the two-dimensional map. \revI{In this pilot study, we focus on using this representation learning approach for similarity searches around the known HSSs. The embedding region in which they live is well-isolated and might harbor additional HSS candidates.}

We retrieved optical information (redshift and classification) for each of the 40 sources (with 75 corresponding observations) in the selected region or island of the embedding space where three of the HSSs' observations cluster. Figure \ref{fig:tsne_zoom} shows a zoom-in of the selected region, where each source is color-coded by redshift. While the vast majority of the objects in the selected regions belong in the Milky Way or nearby galaxies ($z\lesssim0.002$), four objects have redshift $z\gtrsim0.02$. Three of these are known HSSs (two observations of ASASSN-14li), and the fourth source is our newly found TDE, J1035+57, highlighted in Fig.\ \ref{fig:tsne_zoom} by a black empty star, and described at length in Sec. \ref{sec:source}.

\revI{This is a promising result that highlights the applicability and efficiency of ML and representation learning techniques to find and classify X-ray sources of all kinds, not limited to objects exhibiting peculiar short-term variability. An in-depth analysis of all the sources in the selected region, as well as possible interesting neighbors of the HSS falling outside of this cluster, is beyond the scope of this paper, and it will be investigated in a future publication.}

\section{Conclusions} \label{sec:conc}

In this paper, we mined the most recent release of the \cxo\ source catalog searching for HSSs. The study of these peculiar sources allows us to investigate the properties of the less-massive SMBHs and to probe accretion processes on time-scales accessible to human beings. 

Starting from $\approx1.3$~million detections of 407 thousand sources, we selected, owing to their X-ray properties, a parent sample of 125 sources, which we further filtered based on their optical classification, to a final sample of 11 sources. Out of these, six are known HSSs: TDEs, QPEs, and a bonafide accreting IMBH. These six objects are an encouraging confirmation that our filtration process indeed successfully selects HSSs.

We performed a complete X-ray spectral analysis of the remaining five sources and found that one is a misclassified AGN, three are not powered by accretion onto SMBHs, and finally one whose emission is perfectly compatible with being a TDE. This newly found TDE, J1035+57, exploded between late 2001 and early 2002, making it one of the first TDEs ever observed by \cxo. The relatively large mass of the host galaxy, implies that either the SMBH in its center is a low outlier of the $M_\textup{BH}-M_\star$ relation or that this TDE occurred on an IMBH hosted in the same galaxy.

Finally, we explored the applicability of ML and representation learning techniques in identifying HSSs. Exploiting reduced-dimensionality embeddings of part of the \cxo\ catalog, we identified a region of this parameter space within which some known HSSs group together. After retrieving redshift information for the neighboring sources of the clustered HSSs, we noticed that almost all sources are hosted in the Milky Way or in other nearby galaxies, except for one: this source is the newly identified TDE J1035+57. This exciting result proves that representation learning can be employed for X-ray source classification and its power and versatility can be harnessed to identify new HSSs.

\section*{Acknowledgments}
We thank the anonymous referee for insightful comments that improved the paper.
A.S. acknowledges support from a Scholarly Studies Award. PK was supported by NASA through the NASA Hubble Fellowship grant HST-HF2-51534.001-A awarded by the Space Telescope Science Institute, which is operated by the Association of Universities for Research in Astronomy, Incorporated, under NASA contract NAS5-26555. This research has made use of data obtained from the Chandra Data Archive and the Chandra Source Catalog, both provided by the Chandra X-ray Center (CXC). This paper employs a list of Chandra data sets, obtained by the Chandra X-ray Observatory, available at~\dataset[DOI: 10.25574/cdc.334]{https://doi.org/10.25574/cdc.334}.

\vspace{5mm}
\facility{CXO}
\software{Astropy \citep{astropy13,astropy18}, Matplotlib \citep{matplotlib07}, NumPy \citep{numpy20}, Topcat \citep{topcat05}, Ds9}

\appendix

\section{X-ray spectral analysis of excluded sources}\label{app:xray_spec}

In this section, we present the X-ray spectral analysis of the four sources we excluded as their X-ray emission is not compatible with that of an HSS.

\subsection{2XMMi~J215039.5--055335}
This source, located at spectroscopic redshift $z=0.3928$ \citep{connelly12} is not classified as an AGN. It has been observed twice with \cxo, in 2006 (ObsId 6791, P.I. Mulchaey) and ten years later, in 2016 (ObsID 17862, P.I. Lin) for 100~ks and 77~ks of observing time respectively. The source location has also been visited by \xmm, but the analysis of that dataset is beyond the scope of this paper. 

\begin{figure}[t]
\centering
\includegraphics[width = 0.5\textwidth]{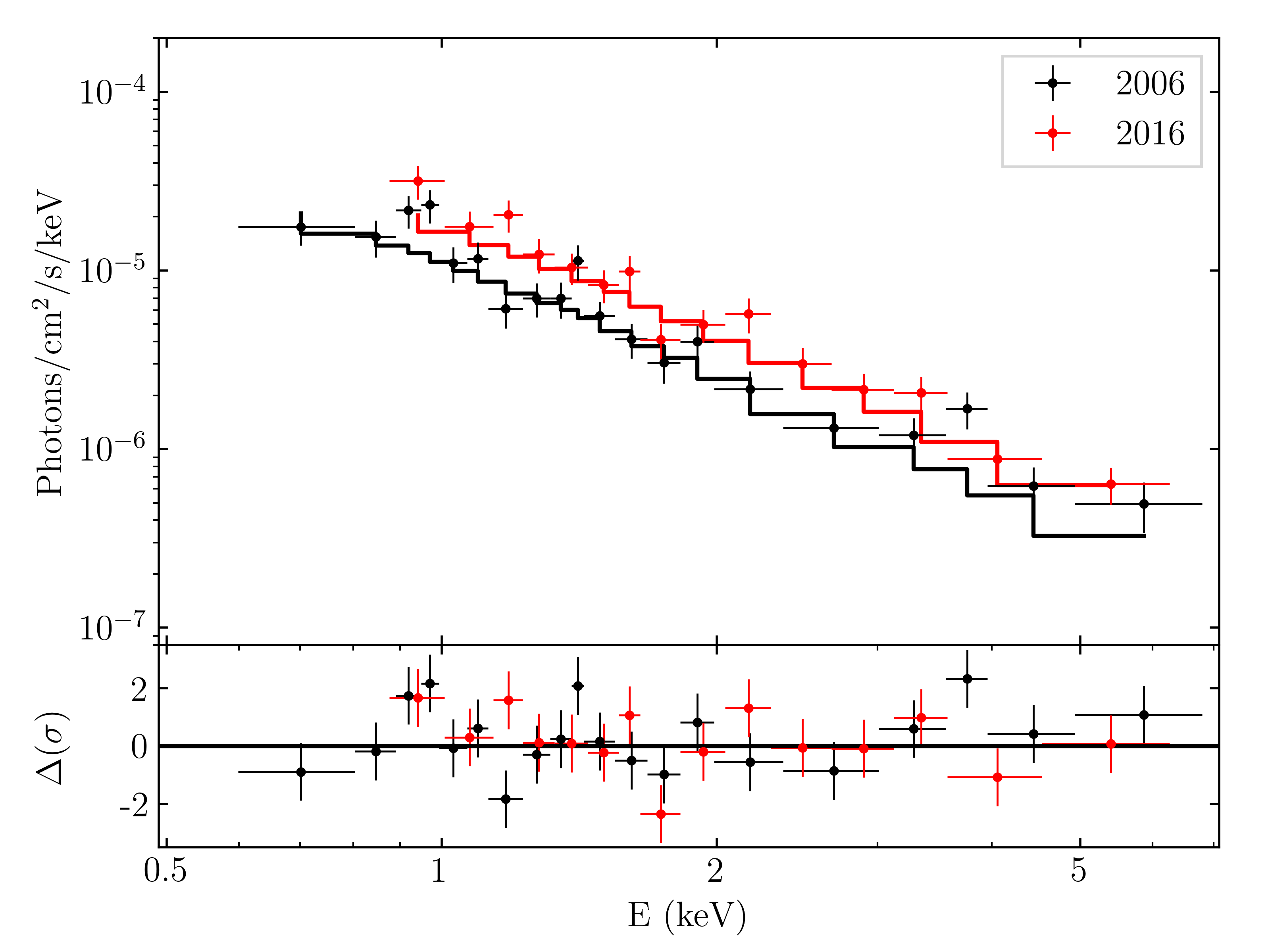}
\caption{X-ray spectrum (upper panels) and residuals (lower panels) taken with \cxo\ in 2006 (in black) and 2016 (in red) of 2XMMi~J215039.5--055335. The solid lines show the best-fitting models described in the text. \label{fig:xagn}}
\end{figure}

The X-ray spectra of the source were regrouped to have at least 20 counts per bin and $\chi^2$ statistic was employed. The spectra were fitted simultaneously and we adopted a simple power-law model, with a neutral absorber with $N_\textup{H}$ fixed at the Galactic value. We obtained an acceptable fit ($\chi^2/\nu=41/31\approx1.3$) with a value of photon index $\Gamma=2.1\pm0.1$ and an unabsorbed luminosity in the 0.5--7~keV band of $(3.8\pm0.2)\times10^{43}$ in 2006 and $(5.5\pm0.2)\times10^{43}$ in 2016. We attempted fitting the spectra with a blackbody model but obtained an unacceptable-quality fit ($\chi^2/\nu=132/31\approx4.2$). The X-ray spectrum in the two \cxo\ visits is reported in Fig. \ref{fig:xagn}. The spectral shape of this source excludes it from being an HSS and coupled with its luminosity suggests that it is in fact a misclassified AGN.

\subsection{IC~1792, 2MASX~J23272553+2635075, ESO~152--37}

\begin{figure*}[t]
\centering
\includegraphics[width = 0.3\textwidth]{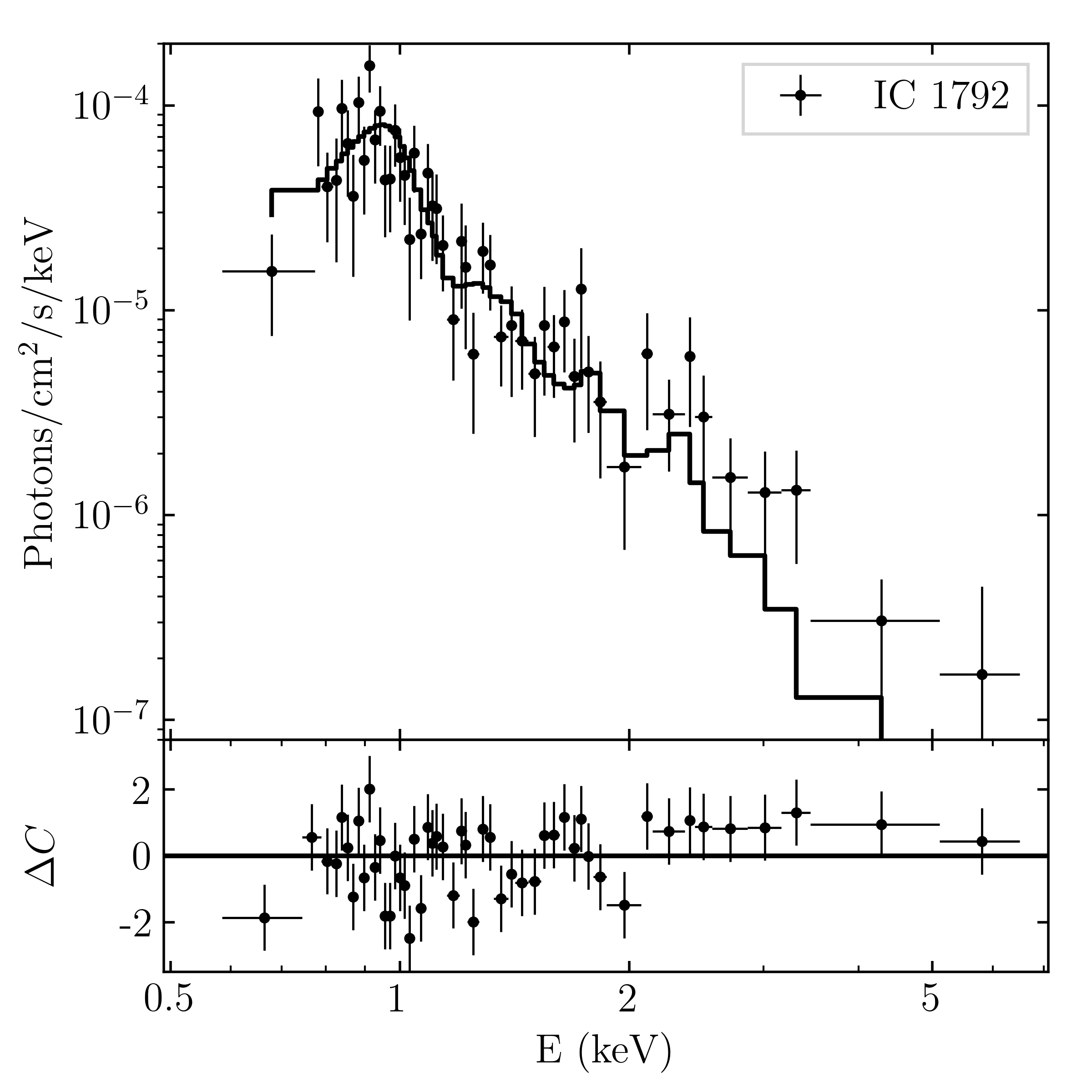}
\includegraphics[width = 0.3\textwidth]{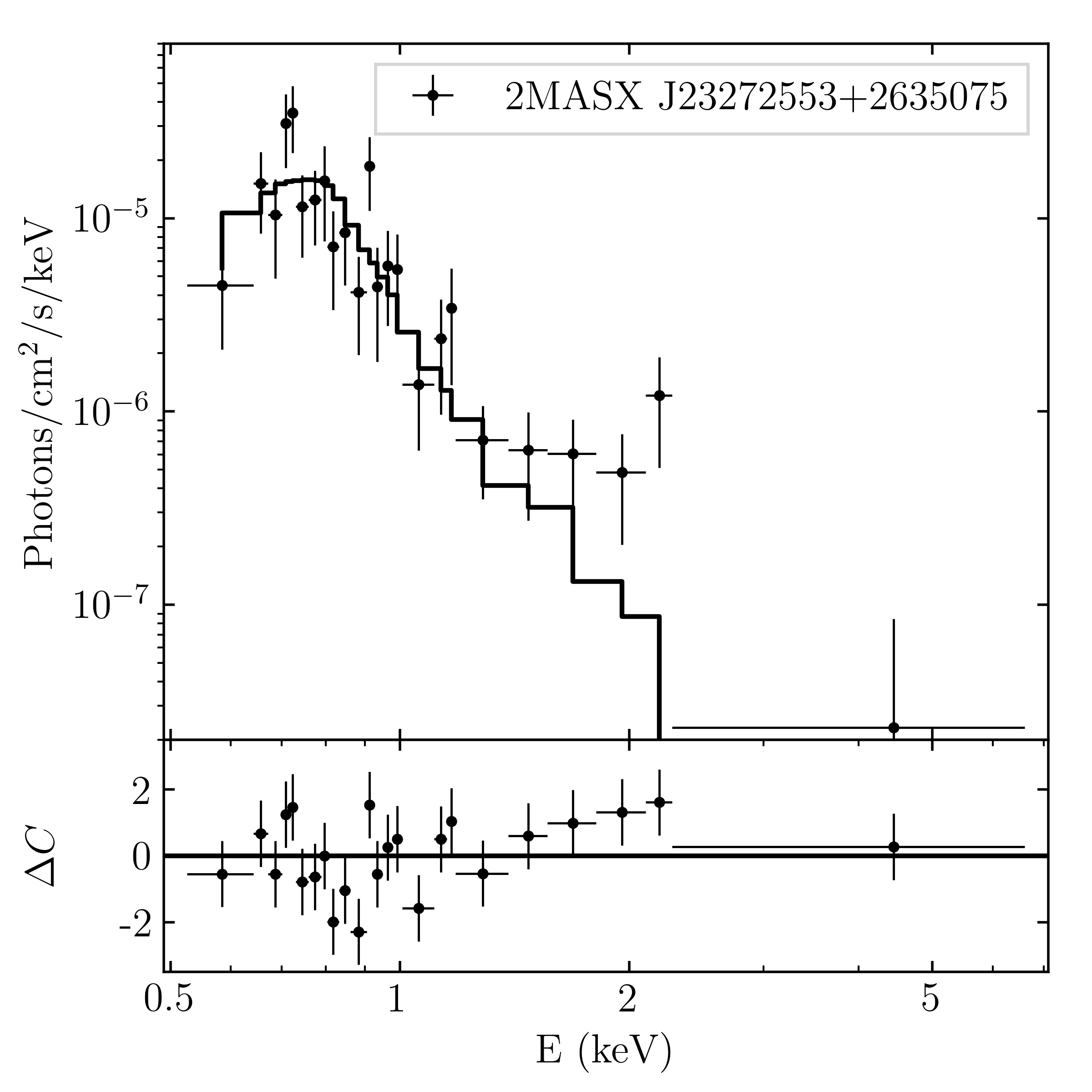}
\includegraphics[width = 0.3\textwidth]{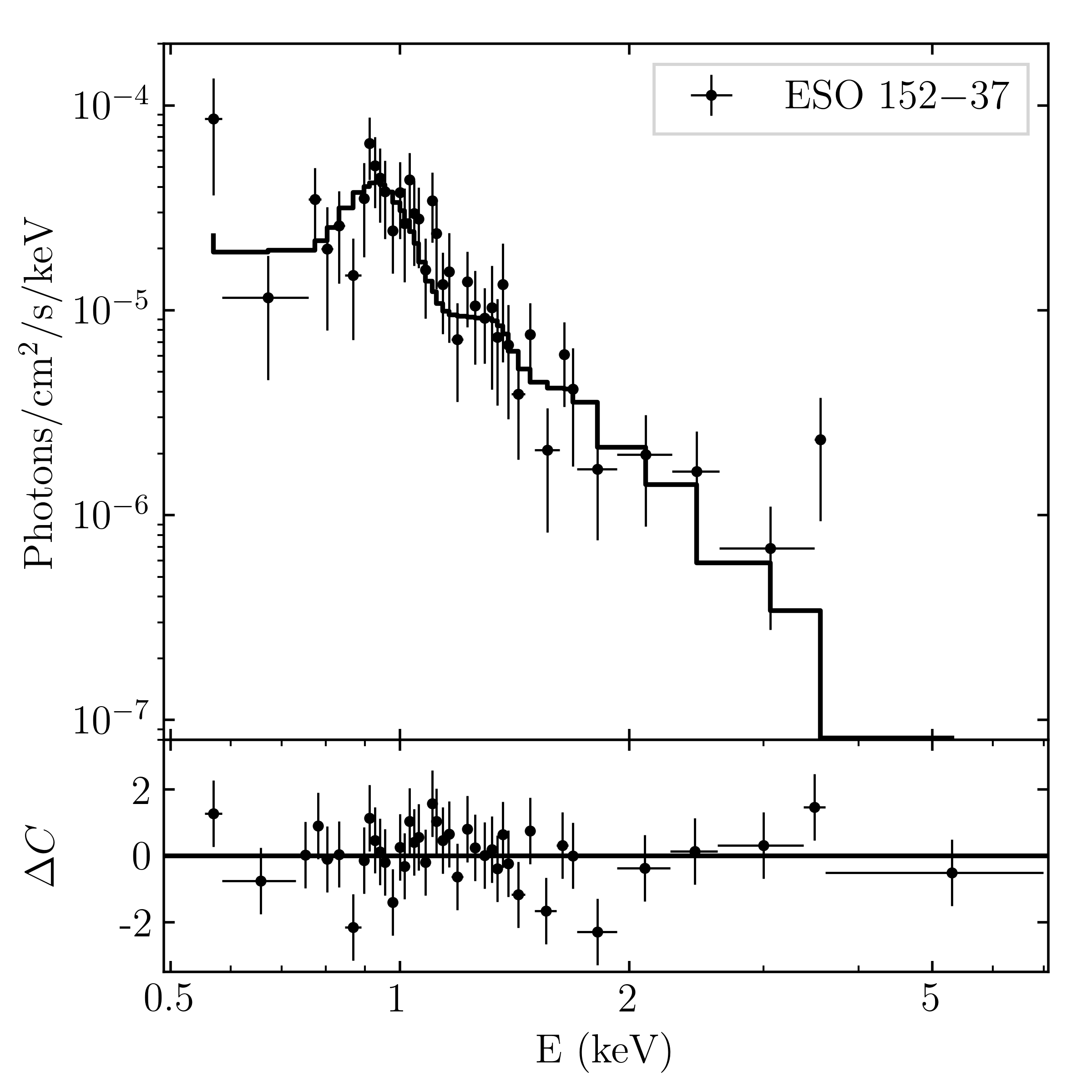}
\caption{X-ray spectra of, from left to right, IC~1792, 2MASX~J23272553+2635075, and ESO~152$-$37. The X-ray spectrum is shown in the upper panels and residuals in the lower ones. The solid lines show the best-fitting models described in the text. \label{fig:xapec}}
\end{figure*}

\revI{These three sources, all classified as quiescent galaxies, are located at redshift $z=0.035$, 0.056, and 0.091 respectively \citep{paturel02,huchra12,jones09}. Each galaxy have observed once with \cxo\, IC~1792 in 2009 (ObsID 11575, P.I. Croston), 2MASX~J23272553+2635075 in 2010 (ObsID 11830, P.I. Garmire), and ESO~152--37 in 2012 (ObsID 13489, P.I. Benson). The X-ray spectra of each source were regrouped to have at least 1 count per bin and modified Cash statistic (W-statistic, in the \texttt{XSPEC} implementation) was employed. We fit the spectra, shown in Fig. \ref{fig:xapec}, with an \texttt{apec} model with a neutral absorber with $N_\textup{H}$ fixed at the Galactic value. For IC~1792 we obtained an acceptable fit ($C/\nu=140/113\approx1.2$) corresponding to a plasma temperature of $kT=1.13\pm0.08$~keV and an unabsorbed luminosity in the 0.5--7~keV band of $(2.0\pm0.1)\times10^{41}$~erg/s. For 2MASX~J23272553+2635075 the best fit ($C/\nu=64.3/60\approx1.07$) corresponds to a plasma temperature of $kT=0.69\pm0.07$~keV and an unabsorbed luminosity in the 0.5--7~keV band of $(7.3\pm0.7)\times10^{40}$~erg/s. Finally, for ESO~152--37 we obtained an acceptable fit ($C/\nu=147.2/171\approx0.86$) with a plasma temperature of $kT=1.42\pm0.09$~keV and an unabsorbed luminosity in the 0.5--7~keV band of $(9.4\pm0.5)\times10^{41}$~erg/s. We attempted fitting each spectrum with a blackbody model but obtained an unacceptable-quality fit or a much-worse-quality fit. The X-ray spectral properties and luminosity of these sources suggest that rather than being powered by ongoing accretion onto SMBHs, their emission is due to circumgalactic gas warming up while falling in the galactic gravitational potential well. }

\bibliography{biblio}{}
\bibliographystyle{aasjournal}

\end{document}